\documentclass{optica-article}

\journal{opticajournal} 

\articletype{Research Article}

\begin{document}

\title{Beam Spliter and Localization Induced by Controlled Perturbations after Time Boundary}

\author{Jing Wang,\authormark{1} Yuetao Chen,\authormark{1} and Shaoyan Gao \authormark{1,2}}

\address{\authormark{1}MOE Key Laboratory for Nonequilibrium Synthesis and Modulation of Condensed Matter, Shaanxi Province Key Laboratory of Quantum Information and Quantum Optoelectronic Devices, School of Physics, Xi'an Jiaotong University, 710049, People's Republic of China.}

\email{\authormark{2}gaosy@xjtu.edu.cn} 


\begin{abstract*} 
The recent investigation into the phenomena of refraction and reflection at temporal boundaries, conducted through the lens of spacetime duality, has attracted considerable scholarly interest. This duality unveils insights into the propagation behaviors of beams at the temporal boundaries of perturbed systems. We have delineated a temporal boundary effect amenable to external control through a specifically tailored driving force, augmented by a time-varying constituent within the driving signal. We then unveil  the phenomenon of beam splitting-both in time refraction and reflection induced by perturbing the lattice’s hopping parameter over time. By introducing varying intensities of aperiodic disorder to the coupling coefficients, we have exercised authority over the reflection angles. Our results lay the groundwork for delving into the temporal evolution of crystalline attributes via temporal boundary effects, while also enabling deliberate manipulation of spatial distribution frequency patterns by regulating the form and magnitude of noise. The results offers a manageable avenue for scrutinizing condensed-matter phenomena through providing an experimentally feasible solution.
\end{abstract*}
\section{Introduction}
Dynamic metamaterials, recognized as fourth-generation metamaterials \cite{8858032}, have garnered significant attention for their ability to manipulate electromagnetic waves through externally driven temporal modulation of material parameters. Unlike static or reconfigurable systems, these materials enable spatiotemporal spectral engineering via intrinsic time-dependent properties, governed by temporal boundaries arising from abrupt parameter changes. Experimental control of time boundary effects has been demonstrated across various platforms, including spatial structures \cite{gratus2021temporal, menendez2017time, maslov2018temporal}, temporal switching in dispersive media \cite{fante1971transmission, fante1973propagation, bakunov2021light, solis2021time, gratus2021temporal}, micromechanical systems \cite{yuan2024above, zhou2020broadband,  arkhipov2024analytical}, and ultra-cold atom lattices \cite{dong2024quantum}. The synthetic frequency dimension, a high-dimensional topological phenomenon, provides a powerful tool for studying temporal boundaries associated with frequency variations \cite{PhysRevResearch.5.L012046}.

Recent advances in time-boundary physics have unveiled novel mechanisms for wave manipulation and topological phenomena in time-modulated systems. Key breakthroughs include enhanced time reflection near critical angles, where sudden permittivity changes transform evanescent fields into propagating pulses \cite{bar2024time}, momentum gap engineering in photonic time crystals (PTCs) for tunable subwavelength amplification \cite{ye2024floquet}, and asymmetric vortex generation in time-switched magneto-optical media \cite{boltasseva2024photonic}. Additionally, Bloch wavefunction overlaps at temporal interfaces have been shown to probe hidden band topology in Chern insulators \cite{wu2024edge}, while atomic-scale light-induced cavities suggest attosecond optical switching capabilities \cite{arkhipov2024analytical}. These developments establish time boundaries as a versatile platform for exploring non-equilibrium wave dynamics and topological phase transitions \cite{yuan2024above, caloz2024emergence}. However, despite extensive research on the robustness of temporal boundary effects and their topological properties \cite{PhysRevResearch.5.L012046, dong2024quantum, PhysRevLett.132.083801}, a gap remains in understanding systems controlled by tunable disorder.

Motivated by space-time duality, time-varying perturbations—conceptually dual to spatial disorder—provide a framework for investigating temporal effects and implementing frequency control in nonstationary systems \cite{Longhi:23, PhysRevB.105.L220202}. While most research has focused on the robustness of temporal boundaries \cite{PhysRevResearch.5.L012046, dong2024quantum, PhysRevLett.132.083801} and their use in classifying Chern insulators \cite{PhysRevLett.132.083801}, the control of frequency transmission via tunable disorder remains unexplored.

In this work, we demonstrate that reflection and refraction at temporal boundaries can be realized in a synthetic frequency dimension system driven by an external field, akin to systems with abrupt coupling constant changes. We observe time-refracted and time-reflected beam splitting under three distinct perturbations and introduce delocalized and localized excitations in the frequency domain following a temporally driven boundary. Furthermore, we explore the relationship between time refraction angles and perturbation amplitudes, offering a new modality for frequency control. Using a synthetic frequency dimension model, we present numerical simulations to validate our findings.

Our study examines the impact of varied perturbations on temporal boundary effects, providing new avenues for frequency manipulation beyond traditional methods \cite{Yuan:16, kumar1990quantum, PhysRevLett.75.2494, Han:21}. This approach also offers a controlled framework for studying condensed matter phenomena, such as optical wave manipulation in engineered media \cite{istrate2006photonic}, distinct from conventional techniques \cite{schwartz2007transport, kuizenga1970fm}.
\section{Synthetic Frequency Dimension Model and External Driving}
\begin{figure}[htbp]
	\centering
	\includegraphics[width=1\textwidth]{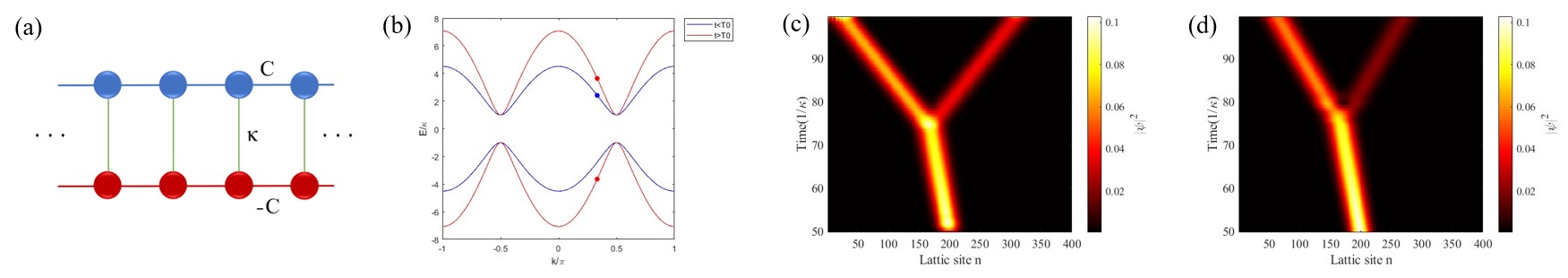} 
	\caption{(a) Schematic representation of the two-leg ladder lattice model. The blue and red spheres represent the sites on the $a$ and $b$ sublattices, respectively. Intra-leg couplings are denoted by $C$ and $-C$ for the $a$ and $b$ sublattices, while the inter-leg coupling is characterized by the parameter $\kappa$. (b) Band structure of the Hamiltonian defined in Eq.~(\ref{eq:4}). The blue curves correspond to the regime before the time boundary ($t < T_0$) with a coupling constant $C = 4$, while the red curves represent the regime after the time boundary ($t > T_0$) with $C = -7$. The dots highlight states on the band structures at the central wave vector $k_0 = \pi/3$. (c) The system consists of $N = 400$ lattice sites arranged into $M = 200$ unit cells, each containing one $a$ site and one $b$ site. Even and odd lattice sites correspond to $a$ and $b$ sites, respectively. At the time boundary $T_0 = 50T$, the coupling constant $C$ undergoes an instantaneous transition from $C = 4$ to $C = -7$. The dynamics are simulated using the matrix exponential method. (d) The system is initialized in the same state as in (c) but evolves according to Eq.~(\ref{eq:6}) with $f(t) = 0$ and $f(t) = 0.3$ for $t > T_0$, respectively.}
	\label{fig:1}
\end{figure}

The Su-Schrieffer-Heeger (SSH) ladder model characterized by an abrupt change describes the temporal boundary  \cite{PhysRevLett.110.203904,PhysRevB.102.161101} while maintaining spatial uniformity. Previous work by Fan et al.\cite{PhysRevResearch.5.L012046} has demonstrated time reflection and refraction by imposing an abrupt change in the coupling strength shown in [Fig.$\ref{fig:1}(a)$] as discribed by the Hamiltonian
\begin{equation}\label{eq:1}
	\begin{aligned}	
	H_{0}&=\sum_{m}[C\hat{a}^{\dagger}_{m}\hat{a}_{m+1}+H.c.]+\sum_{m}[-C\hat{b}^{\dagger}_{m}\hat{b}_{m+1}+H.c.]+\sum_{m}[\kappa \hat{a}^{\dagger}_{m}\hat{b}_{m}+H.c.],
	\end{aligned}
\end{equation}
where the annihilation (creation) operators for particles at sites \textit{a} and \textit{b} of the \textit{m}th unit cell are denoted as $\hat{a}_m$, $\hat{b}_m$ ($\hat{a}_m^\dagger$, $\hat{b}_m^\dagger$), respectively. The intra-leg coupling between adjacent \textit{a} sites is parameterized by $C$, while the corresponding coupling between adjacent \textit{b} sites is set to $-C$,  ensuring an antisymmetric coupling configuration. Additionally, the interleg coupling between sites \textit{a} and \textit{b} within the same unit cell is governed by the coupling constant $\kappa$, as schematically depicted in Fig.$\ref{fig:1}(a)$.

Owing to the translational invariance of the Hamiltonian along the ladder legs, it can be expressed in momentum space as
\begin{equation}
	H_0 = 2C \sum_{k} c^{\dagger}_{k} \mathcal{H}(k) c_k, \label{eq:2}
\end{equation}
where the momentum space annihilation and creation operators are defined as \(c_k^\dagger = \begin{bmatrix} a_k^\dagger \\ b_k^\dagger \end{bmatrix}\) and 
\(c_k = \begin{bmatrix} a_k \\ b_k \end{bmatrix}\) with the momentum space Hamiltonian given by
\begin{equation}
H(k) = \begin{bmatrix}
	2C \cos(k) & \kappa \\
	\kappa & -2C \cos(k)
\end{bmatrix}, \quad \label{eq:3}
\end{equation}
and the dispersion relation as
\begin{equation}
E_{\pm}(k) = \pm\sqrt{(2C \cos(k))^2 + \kappa^2}, \label{eq:4}
\end{equation}
where $E_{+}$ and $E_{-}$ correspond to the upper and lower energy bands[Fig.$\ref{fig:1}(b)$]. The corresponding eigenvectors are expressed as
\begin{subequations}
\label{eq:main}
\begin{equation}
	{\gamma}^{\dagger}_{k} \left| 0 \right>=\frac{1}{N(k)}[\kappa a^{\dagger}(k)+(-2C \ cos \ k + \sqrt{(2C \ cos \ k)^2+\kappa^{2}} b^{\dagger}(k)]\left| 0 \right>,
	\label{eq:5a}
\end{equation}
\begin{equation}
	\tilde{\gamma}^{\dagger}_{k} \left| 0 \right>=\frac{1}{N(k)}[(2C \ cos \ k - \sqrt{(2C \ cos \ k)^2+\kappa^{2}})a^{\dagger}(k)+\kappa b^{\dagger}(k)]\left| 0 \right>,
	\label{eq:5b}
\end{equation}
\end{subequations}
with $\tilde{\gamma}^{\dagger}_{k}$ and $\gamma^{\dagger}_{k}$ representing the new creation operators for the lower ($E_{-}$) and upper ($E_{+}$) energy bands, respectively, and $N(k)$ being the normalization constant. The system inherently ensures momentum conservation across temporal interfaces, as expected due to the preservation of spatial homogeneity.
Temporal evolution of the system initialized in the state described by
\begin{equation}
\psi_{m}(t=0) = \frac{1}{\sqrt{\sigma\sqrt{\pi}}} e^{-\left(\frac{m - m_0}{2\sigma}\right)^2} \begin{bmatrix} 1 \\ 1 \end{bmatrix}, \label{eq:6}
\end{equation}
where $\sigma = 30$, $m_0 = 200$, and $m$ is the index of the lattice site. All initial states discussed in the dynamical evolutions in this paper are prepared in this manner. The results of numerical simulation are depicted in Fig.$\ref{fig:1}(c)$. The time-refracted beam, propagating in the same direction as the incident beam, and the time-reflected beam, propagating in the opposite direction, are clearly distinguishable following the temporal boundary at $t=T_0$.

We denote the simple notation of the energy bands as $\epsilon_{k} = E_{+}(k)$, corresponding to the Hamiltonian
\begin{equation}
H_{0} = \sum_{k} [\epsilon_{k}\gamma_{k}^{\dagger}\gamma_{k} - \epsilon_{k}\tilde{\gamma}_{k}^{\dagger}\tilde{\gamma}_{k}]. \nonumber
\end{equation}

If we consider a property operator $\hat{P}$, the system exhibits chiral symmetry, expressed as $\{H_0,\hat{P}\}=0$. To facilitate our discussion, we now define the operator $\hat{P}$ as $\hat{P} = \sum_{k} \hat{P}_{k}$, where $\hat{P}_{k}$ is represented in the eigenbasis as $\hat{P}_{k} = \tilde{\gamma}^{\dagger}_{k}\gamma_{k} + \gamma_{k}^{\dagger}\tilde{\gamma}_{k}$, and within each $k$ subspace, $\hat{P}_{k}\gamma_{k}^{\dagger}|0\rangle = \tilde{\gamma}_{k}^{\dagger}|0\rangle$. The operator $\hat{P}_k$ induces coupling between the paired states $|\gamma_{k}\rangle$ and $|\tilde{\gamma}_{k}\rangle$ with the same $k$.

The total Hamiltonian is given by
\begin{equation}
\mathcal{H} = H_0 + f(t)\hat{P}. \label{eq:7}
\end{equation}
We observe that manipulating the functional form of $f(t)$ can induce a variety of novel phenomena in the temporal boundary dynamics of light, as illustrated in Fig.$\ref{fig:1}(d)$. For $t > T_0$, following the temporal boundary, the driving term $f(t)\hat{P}$ excites paired states that propagate in opposite directions within the frequency domain. These states are identified as the time-refracted beam and the time-reflected beam, respectively, and their properties are governed by the specific form of $f(t)$.

In subsequent sections, we employ numerical simulations to elucidate the mechanism of light reflection at temporal boundaries within a two-site ladder model. Furthermore, we explore novel phenomena arising from the introduction of various forms of disorder into the system, including both static and dynamic perturbations, to understand their impact on wave propagation and localization.

\section{Numerical Results}

In our investigation, we have identified a significant consequence arising from the inherent central symmetry of the Hamiltonian's band structure: the substitution of $k$  with $-k$ in Eq.$(\ref{eq:5b})$ leaves the expression for $\tilde{\gamma}_{k}$ effectively unchanged. This invariance plays a critical role in understanding the function of the operator $\hat{P}$, which, when applied, establishes a connection between the energy states $\epsilon_{k}$ and $-\epsilon_{-k}$, thereby acting as a key mechanism in the system's dynamics. This symmetry-driven behavior underpins the robustness of the observed phenomena and provides a foundation for further exploration of topological and dynamical properties in similar systems.

We introduce a novel operator configuration denoted as $\hat{P}_0$. By substituting $k $ with $ -k $ in the operator $ \tilde{\gamma}_{k}^{\dagger}$, we obtain 
	\begin{align*}
	\tilde{\gamma}^{\dagger}_{k} |0\rangle &= \frac{1}{N(k)} \left[ (2C \cos(-k) - \sqrt{(2C \cos(-k))^2 + \kappa^2}) a^{\dagger}(-k) + \kappa b^{\dagger}(-k) \right] |0\rangle \\
	&= \frac{1}{N(k)} \left[ (2C \cos(k) - \sqrt{(2C \cos(k))^2 + \kappa^2}) a^{\dagger}(-k) + \kappa b^{\dagger}(-k) \right] |0\rangle,
	\end{align*}
which corresponds to the energy $ E_{-}(-k)$ and demonstrates central symmetry in momentum space regarding the eigenvector $\gamma^{\dagger}_{k} $ corresponding to $ E_{+}(k) $.

Incorporating a time-dependent, random driving term in the form of$ f(t)\hat{P}_0 $ into the system's Hamiltonian, 
\begin{equation}
\mathcal{H}^{'}=H_0+f(t)\hat{P_0},\label{eq:8} 
\end{equation}
endows it with a new dimension of dynamical complexity. The operator $ \hat{P}_k $introduces a coupling between the paired states $ |\gamma_{k}\rangle $ and $|\tilde{\gamma}_{k}\rangle$ with identical absolute values of $|k|$,  which significantly distinguishes it from the operator $ \hat{P} $.  This intentional inclusion triggers a cascade of quantum mechanical responses that are deeply influenced by the underlying symmetry of the system. From this perspective, we can explore the intricate interplay between disorder, temporal driving, and symmetry, potentially revealing previously unexplored quantum phenomena. Such phenomena may include novel topological phases or symmetry-protected dynamical behaviors, offering new insights into the role of symmetry in driven quantum systems.

We observe that introducing time-dependent random driving into Eq.($\ref{eq:8}$) with $f(t) = \delta h R(t)$, or a specific modulation of the coupling coefficient $\kappa$, results in the emergence of two distinct temporally separated refracted and reflected waves beyond the temporal boundary. Three distinct mechanisms are identified as responsible for these excitations at the temporal boundary: Firstly, the system described by Eq.$(\ref{eq:8})$ experiences perturbations in the strength of the external driving. Secondly, a temporally varying perturbation-induced coupling modulation of the form
\begin{equation} 
\kappa(t) = \kappa + \delta h R(t), \label{eq:9} 
\end{equation} 
gives rise to additional band excitations. Thirdly, a perturbation of the coupling constant along the synthetic frequency dimension, expressed as 
\begin{equation} 
\kappa(m) = \kappa + \delta h R(m). \label{eq:10} 
\end{equation}.

The incorporation of varying strengths of \(\delta h\) and different forms of \(R(x)\) (where \(x = t\) or \(m\)) in our system confirms the possibility of achieving split refracted and reflected beams at specific parameter values. Additionally, we find that time-varying disordered external drives can control the localization and delocalization of the temporal structure's frequency distribution, offering a powerful tool for manipulating wave dynamics. By introducing non-periodic sequences with distinct modulation strengths into \(\kappa\), we observe that within a certain range of \(\delta h\), delocalization may be suppressed, while the angles of the refracted and reflected beams can be precisely tuned. This tunability provides a pathway for designing advanced optical devices with tailored beam-steering capabilities.

Our numerical simulations investigate ordered to disordered control methods via \( R(x) \) and \( \delta h \):  
(a) \( R(x) = \cos(x) \), with \( \hat{U}(R(x)) \) determined by \( H(R(x)) \);  
(b) Thue-Morse sequences \( R(x) = \text{TM}_{\text{sequence}}(x) \) \cite{PhysRevLett.127.050602}, defining \( H_{\pm} = H(R(x=0,1)) \) and unitary operators \( U_{\pm} = \exp(-ixH_{\pm}) \), with recursive relations \( U_{x+1} = \tilde{U}_{x}U_{x} \) and \( \tilde{U}_{x+1} = U_{x}\tilde{U}_{x} \);  
(c) Random disorder \( R(x) = \text{rand}(x) \) \cite{mafi2015transverse}, uniformly distributed in \([-0.5, 0.5]\).  

We analyze frequency-domain deformations of time-refracted and time-reflected beams using \( \hat{U}(t) \) for arbitrary \( f(t) \) and \( \kappa \).

\subsection{Three approaches Inducing Splitting of Refracted and Reflected Beams}
Our numerical simulations demonstrate that introducing regular, finite-range interventions enables the splitting of refracted and reflected beams, providing a robust mechanism for controlling wave dynamics in temporally modulated systems.

\begin{figure}[htbp]
\centering
\includegraphics[width=0.8\textwidth]{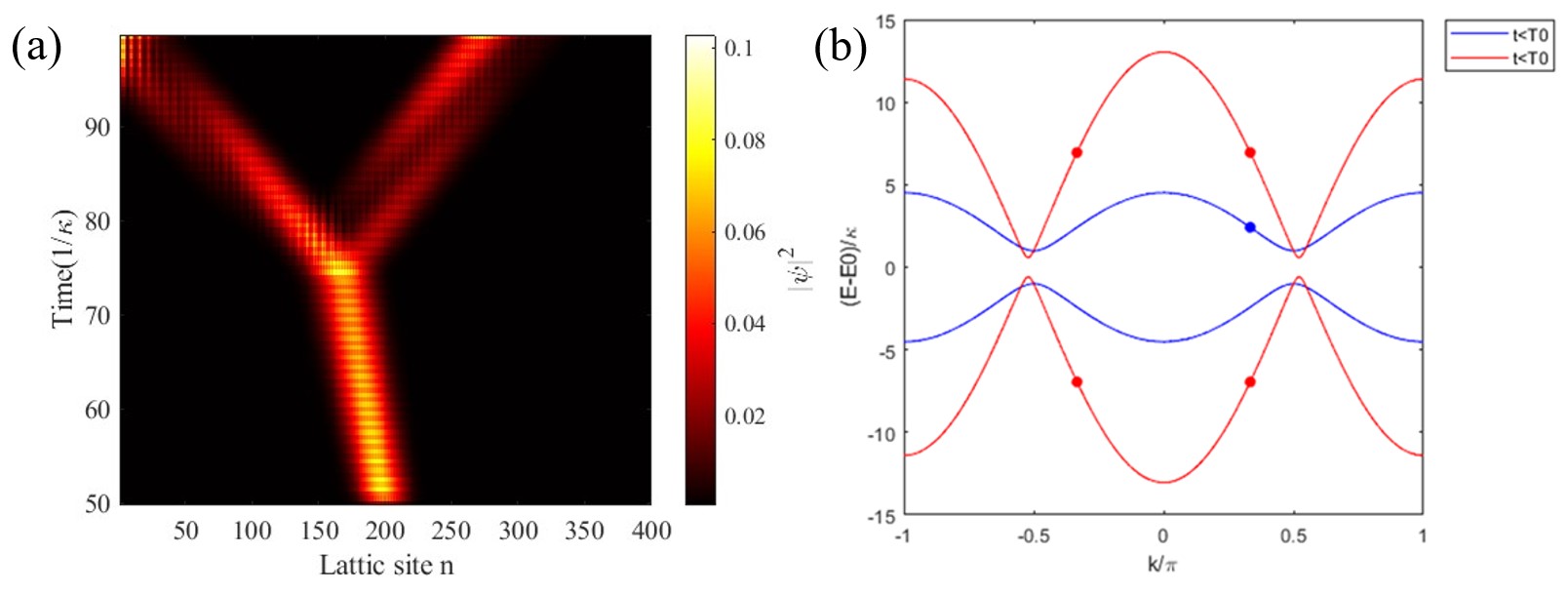} 
\caption{(a)Temporal evolution according to Eq.($\ref{eq:6}$), with $f(t) = 0$ and $f(t) = 0.6 \  TM_{sequency}(t)$ after the time boundary $T_0 = 50T$, (b) Band structure of the Hamiltonian given by Eq.($\ref{eq:8}$), where blue indicates the state before the time boundary ($t < T_0$) with a coupling constant $f(t) = 0$, while red indicates the state after the time boundary ($t > T_0$) with $f(t) = 0.6\ TM_{sequency}(t)$. Dots represent states on the band structures with a central wave-vector $k_0 = \pi/3$, respectively.} 
\label{fig:2} 
\end{figure}

\begin{figure}[htbp] 
\centering
\includegraphics[width=0.8\textwidth]{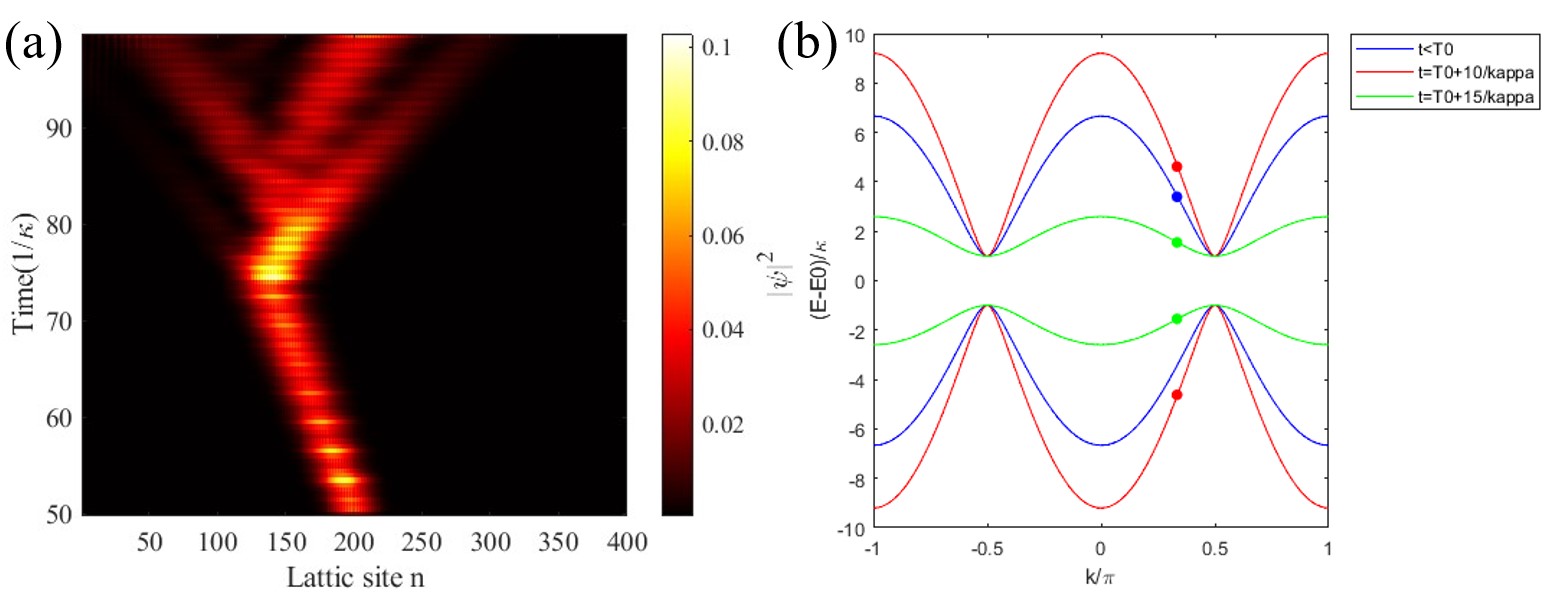} 
\caption{Temporal Evolution and Hamiltonian Band Structure (a) Temporal evolution of the system with a cosine-modulated coupling coefficient $\kappa=6 \cos(t)$. (b) Hamiltonian band structure at different time instances: $t<T_0$ (blue line), $t=T_0+10T$ (red line), and $t=T_0+15T$ (green line). The dots denote states on the band structure corresponding to the central wave vector $k_0 = \frac{\pi}{3}$.} 
\label{fig:3}
\end{figure}

In quantum optics, external driving forces in ladder systems reveal intriguing light-matter interactions. We analyze the Hamiltonian in Eq.~(\ref{eq:9}), where the driving amplitude \( R(t) = \text{TM}_{\text{sequence}}(t) \cdot \delta h \) (with \( 0 < \delta h < 1 \)) induces a bifurcation of frequency-refracted and frequency-reflected beams (Fig.~\ref{fig:3}(a)).

This bifurcation stems from symmetric coupling around the central energy band (Fig.~\ref{fig:2}(b)), mediated by the driving operator \( \hat{P}_0 \) in Eq.~(\ref{eq:8}). At the temporal boundary, transitions from pre-boundary (blue line/dots) to post-boundary (red line/dots) states cause beam splitting for \( t > T_0 \). The non-uniform energy distribution arises from \( f(t) \)-modulated transitions between the four frequencies (red dots).

Our study of light dynamics in periodically driven systems reveals a secondary effect explaining frequency beam splitting [Fig.~\ref{fig:3}(a)]. This stems from time-varying cosine modulation of \( \kappa \) in the undriven Hamiltonian (Eq.~(\ref{eq:9})). Fig.~\ref{fig:3}(b) shows the splitting: red dots mark central beams, while green dots denote outer beams.

The pattern in Fig.~\ref{fig:3}(a) arises from discrete band excitations driven by \( f(t) \), influencing light trajectories. This discovery advances photonic band structure theory and enables new optical engineering applications.

\begin{figure}[htbp]
\centering
\includegraphics[width=0.8\textwidth]{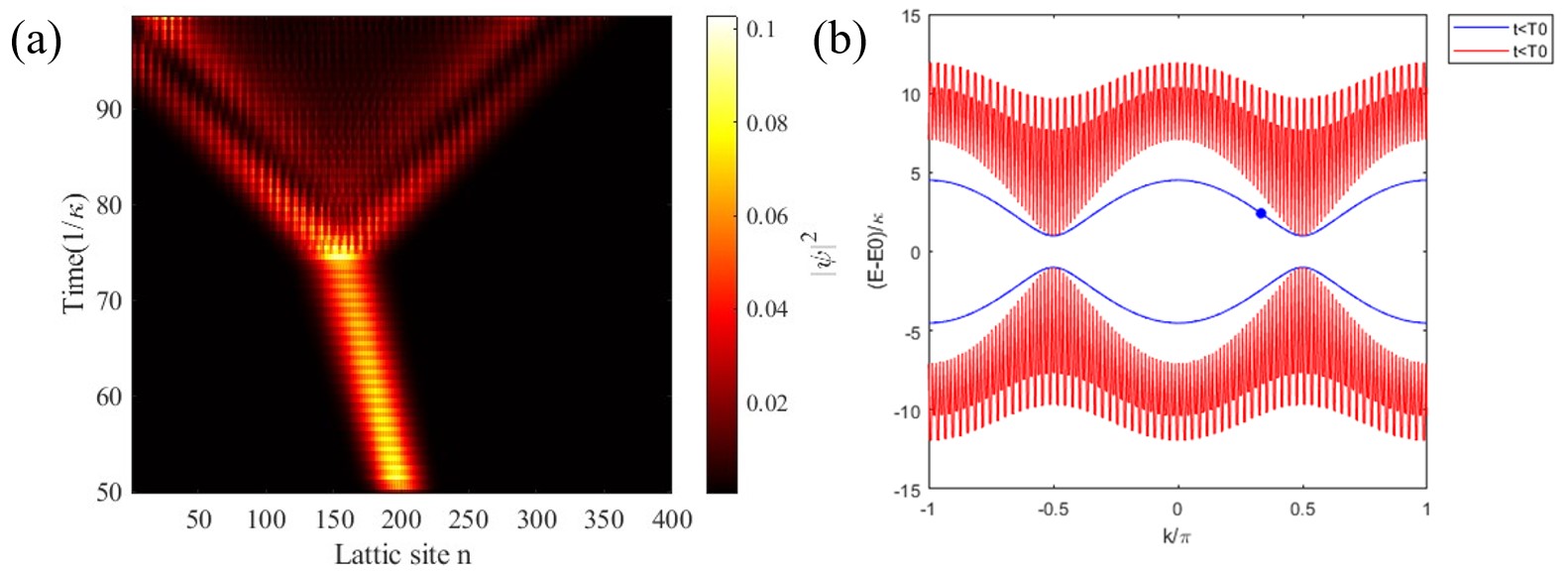} 
\caption{Temporal evolution of the system (a) the coupling parameter $\kappa$ is modulated by a cosine function, $\kappa=0.5\cos(n)$, along the lattice direction. Panel (b) displays the band structure without modulation in blue, while the modulated band structure is depicted in red. The dots represent the states on the unmodulated band at the central wave vector $k_0 = \frac{\pi}{3}$.} 
\label{fig:4}
\end{figure}
We introduce a temporally modulated cosine function to \( \kappa \) (Fig.~\ref{fig:4}), inducing a distinct bifurcation in refracted and reflected beams. This arises from the interplay between low-frequency refraction and high-frequency band modulation (Fig.~\ref{fig:4}(b)), demonstrating precise control over lattice parameters and enabling advances in photonic systems.

\subsection{Localization of Frequency Distribution Induced by Time-Varying Disordered External Drives}
\begin{figure}[htbp]
\centering
\includegraphics[width=1.0\textwidth]{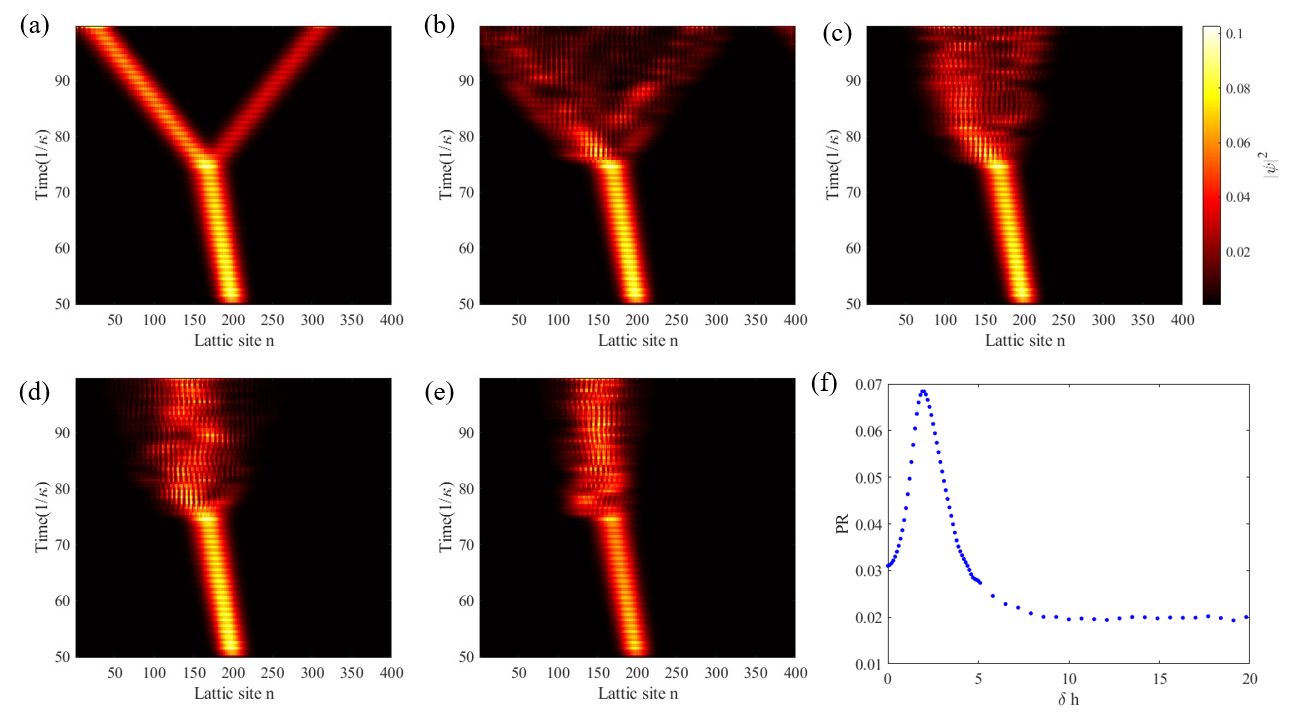} 
\caption{Temporal evolution of eigenstates governed by the Hamiltonian given in Eq.($\ref{eq:8}$) at $t = 70T$, while the driving profile is $f(t) = \delta h \ \text{rand}(T)$. The amplitudes $\delta h$ considered are (a) $\delta h = 0$, (b) $ \delta h = 1$, (c) $\delta h = 1.8$, (d) $\delta h = 8$, and (e) $ \delta h = 10 $. Moreover, (f) presents the average participation ratio(PR) calculated for various  $\delta h$ values, averaging over 100 realizations of disorder.}  
\label{fig:5}  
\end{figure} 

We introduce a time-dependent external drive $f(t)\hat{P}$, where $f(t) = \delta h \cdot \mathrm{rand}(t)$ represents a disorder potential that updates its value randomly at each time step after the temporal boundary $T_0$. The random term $\mathrm{rand}(t)$ follows a distribution with zero mean and unit variance, scaled by the amplitude $\delta h$. For weak disorder ($\delta h \ll 1$), the frequency distribution largely preserves features of the unperturbed system, maintaining distinct temporal refraction/reflection beams [Figs.~\ref{fig:5}(b)--(c)].As $\delta h$ increases, intermediate frequency components emerge between the original refraction and reflection bands, indicating partial delocalization. Further enhancement of $\delta h$ gradually suppresses the excitation bandwidth, causing the temporal refraction/reflection beams to converge as spectral boundaries [Fig.~\ref{fig:5}(d)]. 
We analyze these patterns using the average participation ratio (PR), \( \overline{R}/L \) \cite{kramer1993localization, 10.21468/SciPostPhys.4.5.025}, where \( \overline{R} = \langle R_t \rangle_{t,r} \) and \( R_t = 1/\sum_i |\psi^t_i|^4 \), with \( \psi^t_i \) representing the eigenfunction value at position \( i \) and time step \( t \). Beyond a critical value $\delta h_c = 1.8$, the frequency distribution reaches maximum delocalization [Fig.~\ref{fig:5}(f)], corresponding to the full excitation of extended modes. For $\delta h > 8.6$, the system enters a saturation regime where the frequency distribution becomes amplitude-independent, achieving disorder-controlled spectral localization under time-modulated drives.

Band structure analysis reveals that the time-step disorder induces stochastic inter-band transitions, primarily between the red bands shown in Fig.~\ref{fig:1}(b). This explains the observed spectral broadening at moderate $\delta h$. Stronger disorder suppresses both band transitions and spectral extension, as evidenced by the merging of upper/lower bands. Ultimately, high disorder amplitudes ($\delta h > 8.6$) completely inhibit band-related transitions, suppressing temporal refraction/reflection effects while stabilizing frequency localization. 

\subsection{Changes in Time Refraction Angle Induced by Disorder}
\begin{figure}[htbp] 
\centering
\includegraphics[width=1\textwidth]{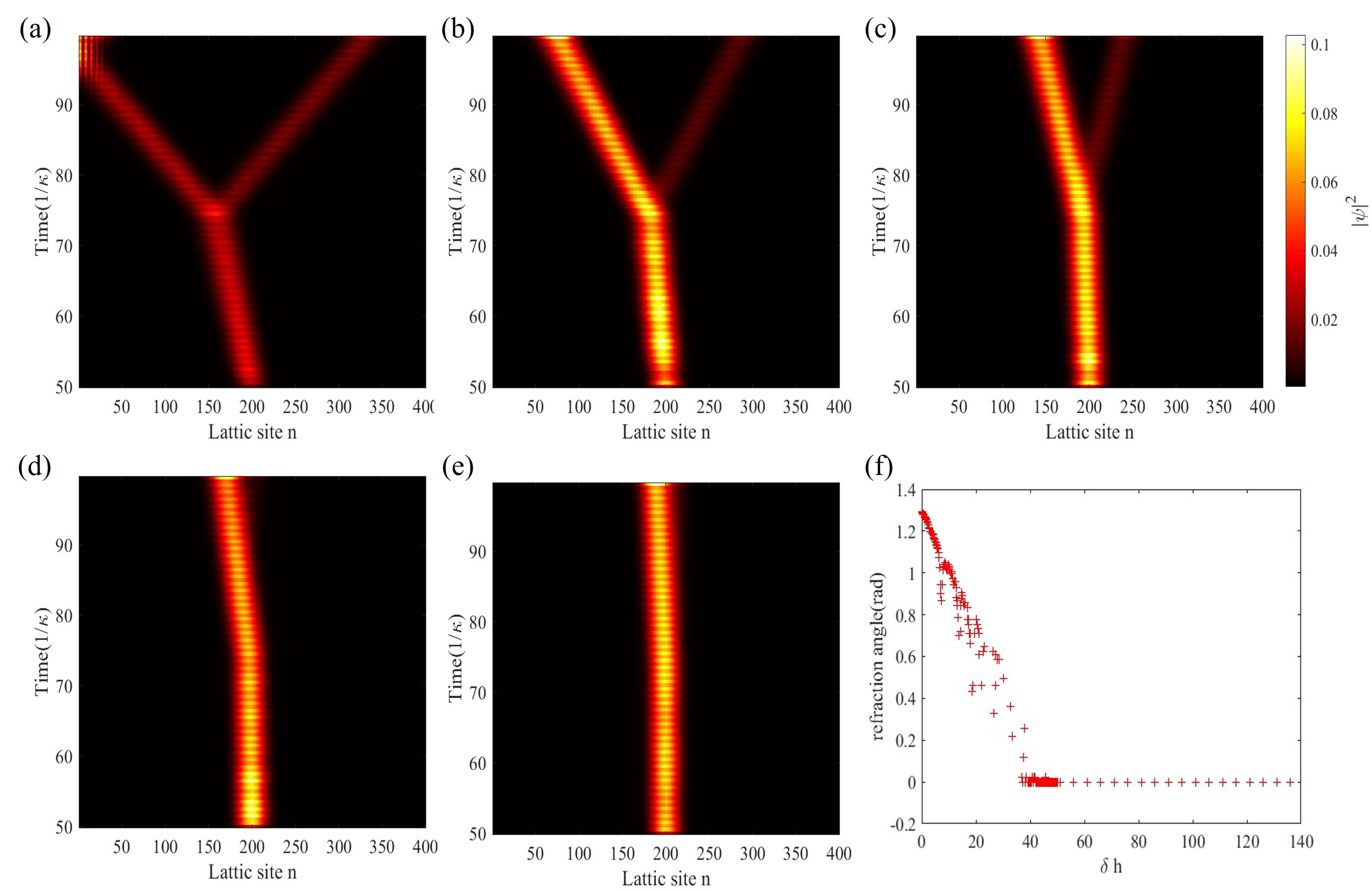} 
\caption{Temporal evolution of eigenstates according to the Hamiltonian $H_0$ at different times, while the coupling coefficient $\kappa(t)$ is given as $\delta h \ TM_{sequence}(T)$. The amplitudes $\delta h$ at (a) $ \delta h = 0$, (b) $\delta h = 5$, (c) $\delta h = 20$, (d) $\delta h = 40$, and (e) $\delta h = 50$ are shown. Additionally, (f) displays the calculated time refraction angles for various values of $\delta h$.} 
\label{fig:6}
\end{figure}

The decrease in the temporal refraction angle is linked to the sharpness of the step function at the temporal boundary, as established in \cite{dong2024quantum}. By introducing a time-dependent Thue-Morse sequence quasi-periodic modulation to the coupling coefficient $\kappa$, we demonstrate precise control over the refraction angle of temporal-refracted beams without inducing wave packet delocalization. Figs.~\ref{fig:6}(a)--(f) illustrate the gradual reduction of temporal refraction angles with increasing disorder strength $\delta h$ through time-evolving patterns.
Fig.~\ref{fig:6}(g) reveals the quantitative relationship between disorder strength and refraction angle: The refraction angle decreases linearly with $\delta h$ when $\delta h < 35$, and vanishes completely at $\delta h = 35$ where both temporal refraction and reflection effects disappear. This critical behavior is shown in Fig.~\ref{fig:6}(f), where the system exhibits neither frequency shift in incident beams nor post-boundary refraction/reflection. The suppression mechanism originates from the Thue-Morse sequence modulated $\kappa(t)$, which avoids multi-band transitions while smoothing the energy bands at temporal boundaries. This band flattening progressively inhibits group velocity dispersion, and ultimately reduces the group velocity to zero at $\delta h = 35$, freezing momentum evolution and suppressing temporal boundary effects.

Furthermore, increasing $\delta h$ enhances the spectral weight of refracted frequencies. This stems from three intertwined mechanisms: (1) Disorder-induced broadening of the temporal boundary step function weakens the abrupt parameter switching; (2) Energy redistribution preferentially enhances refracted components near critical angles, contrasting with dielectric-induced energy transfer in temporal boundaries \cite{bar2024time}; (3) Disorder suppresses inter-band transitions at temporal boundaries, favoring upper-band frequency retention over lower-band occupation. After multiple disorder-scattering events, the system develops dominant spectral components aligned with the initial group velocity direction. These effects collectively demonstrate that the switching rate at temporal boundaries governs the refraction-reflection distribution.

\section{Conclusions and discussions}
In summary, we have demonstrated time refraction and reflection effects without altering the system's coupling parameters, using dynamic driving over a finite duration to introduce a novel approach for temporal boundaries. We identified three mechanisms driving the bifurcation of time-reflected beams, expanding the toolkit for frequency manipulation. By introducing specific driving forms into temporally bounded systems, we coupled excitations across wave vectors, observing both delocalized and localized excitations through modulated oscillatory intensities. Furthermore, we achieved controlled variation of time refraction and reflection angles by incorporating time-varying disorder into the coupling parameters of an undriven system.

These findings provide insights into controllable light manipulation in synthetic frequency dimensions and pave the way for exploring many-body effects \cite{abanin2017recent}, non-equilibrium dynamics at temporal scales \cite{zhang2017observation, xu2021realizing}, and novel phenomena in time-varying crystals \cite{PhysRevLett.126.163902}. The implications for topological phenomena \cite{PhysRevLett.132.083801, Lustig:18} and optical metafaces \cite{su2018advances} are particularly intriguing.
\bibliography{sample}
\end{document}